\documentclass[aps,prl,twocolumn,showpacs]{revtex4}

\usepackage{dcolumn}
\usepackage{amsmath}
\usepackage{amssymb}
\usepackage{graphicx}

\begin{document}

\newcommand \be {\begin{equation}}
\newcommand \ee {\end{equation}}
\newcommand \bea {\begin{eqnarray}}
\newcommand \eea {\end{eqnarray}}
\newcommand \nn {\nonumber}
\newcommand \ve {\varepsilon}

\title{Designer Patterns: Flexible Control of Precipitation through
Electric Currents}

\author{I. Bena$^{1}$, M. Droz$^{1}$, I. Lagzi$^{2}$, K. Martens$^{1}$, Z. R\'acz$^{3}$, A. Volford$^{4}$}
\affiliation{${}^1$Theoretical Physics Department,
University of Geneva, CH-1211 Geneva 4, Switzerland}
\affiliation{${}^2$Institute of Chemistry, E\"otv\"os University, 1117 Budapest, Hungary}
\affiliation{${}^3$Institute for Theoretical Physics--HAS,
E\"otv\"os University, 1117 Budapest, Hungary}
\affiliation{${}^4$Department of Physics, University of Technology and Economics, 1521 Budapest, Hungary}

\date{\today}

\begin{abstract}
Precipitation patterns generated by  $A^{+}+B^{-}\to C$ type
reaction-diffusion processes are studied. It is shown both
theoretically and experimentally that the patterns can be controlled
by an appropriately designed, time-dependent electric current
in the system. We describe examples of current dynamics yielding
periodic bands of prescribed wavelength, as well as more complicated
structures. The pattern control is demonstrated experimentally on the
reaction-diffusion-precipitation process ${\rm {2AgNO_3+K_2Cr_2O_7}}$ $ \to
\underline{{\rm {Ag_2Cr_2O_7}}}$ $+ {\rm {2KNO_3}}$ taking place in a gel.

\end{abstract}


\maketitle

Spontaneous pattern formation can be observed
at all lengthscales~\cite{shinbrot} and much effort has been devoted
to gaining insight into the dynamics of these
processes~\cite{CrossHoh}. One of the aims of these studies
is to reproduce and
control the emerging patterns, thereby opening possibilities
for technological applications such as downsizing  electronic
devices~\cite{lu}. We focus here on an important class of reaction-diffusion 
systems yielding precipitation patterns~\cite{Henisch}. Since these patterns
emerge in the bulk, they have been studied recently in connection
with the so-called {\it bottom-up} (bulk) design methods, as
contrasted to {\it top-down} ones where material is removed
to create structures (as e.g. in case of lithography).
It has been found that the precipitation patterns can be
influenced by appropriately chosen
geometry~\cite{giraldo}, boundary conditions~\cite{grzybowski},
or by a combined tuning of the initial and 
boundary conditions~\cite{tsapatsis,Guiding-field}. 
The above control methods are straightforward but unwieldy,
and more flexible approaches are clearly needed.
Here we introduce a novel method  based on the use of
pre-designed electric currents for regulating the dynamics of the
reaction zones in the system.
The power of this method is verified experimentally
by producing periodic precipitation
patterns with controlled spacing, as well as more complex structures.

Our idea stems from the observation that precipitation
patterns are often formed in the wake of moving reaction
fronts~\cite{CrossHoh,Henisch}.
The motion and reaction dynamics of the front determine where and when
the concentration of the
reaction products crosses threshold levels thus inducing
precipitation. Consequently, and this is the essence of our proposal,
control over the precipitation
pattern can be realized through {\em controlling the properties
of the reaction front}.
In order to explain how this can be done, we turn to the concrete
example of Liesegang patterns~\cite{Liese-1896,Henisch}.

\begin{figure}[t!]
\includegraphics[width=0.9\columnwidth,clip]{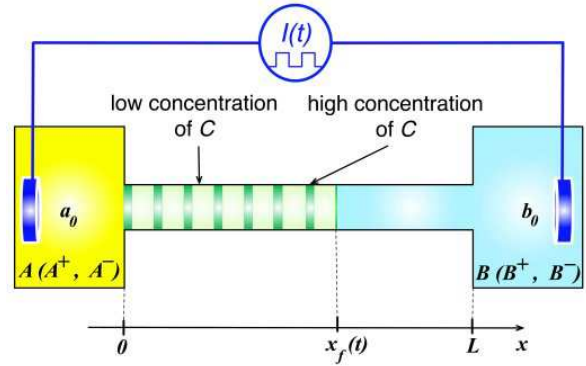}
\caption{Experimental setup for producing Liesegang precipitation patterns
as described in the text.
The controlling agent is the generator providing electric current
$I(t)$ with a prescribed time-dependence.}
\label{figure1}
\end{figure}

%%%%%%%%%%%%%%%%%%%%%%%%%%%%%%%%%%%%%%%%%%%%%%%%%%%%%%%%%%%%%%%%%%%%%%%
In a somewhat generic description, the Liesegang
dynamics consists of the reaction $A^++B^-\to C$
of the ions of two electrolytes $A\equiv (A^+,A^-)$
and $B\equiv (B^-,B^+)$, followed by the precipitation of
the reaction product $C$. The electrolytes are initially separated
(see Fig.~\ref{figure1} for a typical experimental setup) with
the inner electrolyte $B$ homogeneously dissolved in the gel
inside a tube. The
outer electrolyte $A$ is in an aqueous solution which is
brought into contact with the
gel at the start of the experiment. The initial concentration $a_0$
of $A$ is much larger than that $b_0$ of $B$ and, consequently,
$A$ invades the gel and a {\em reaction front} forms and
moves along the tube. 

The relevant properties of this front in the absence of 
an electric field are simple and well-known~\cite{GR1988,MatPack98}:\\
(i) The width of the front is practically negligible.\\
(ii) The front moves diffusively, i.e., its position along the tube
is given by $x_f(t)=\sqrt{2D_ft}$,
with the diffusion coefficient $D_f$ determined
by $a_0$, $b_0$, and the
diffusion coefficients of the reagents.\\
(iii) The concentration $c$ of the reaction product $C$ left in the wake
of the front is constant, $c=c_0$. The value of $c_0$ depends on $a_0$,
$b_0$, the
diffusion coefficients, and on the rate
$k$ of the reaction $A^++B^-\overset{k}{\rightarrow} C$.

The second step in the pattern formation, namely the
phase separation of the $C$-s, takes place only if their
local concentration $c$ is above a precipitation
threshold, $c>c^*$. The experimental setup is thus chosen so that
$c_0>c^*$. The precipitation pattern itself is then the result of a
complex interplay of the production of $C$-s by the front and the
ensuing phase separation dynamics in the wake of the front. Namely, the front
produces a precipitation band at the beginning since $c_0>c^*$ just
behind the front. This band acts as a sink for the newly produced $C$-s
and thus their concentration in the front decreases below $c^*$.
As the front moves far from the existing band,  the
depletion effect diminishes and the $c$ in the front
can again exceed $c^*$ thus leading to the formation of
the next band. A quasiperiodic reiteration of the above process
yields the Liesegang patterns (lowest panel in Fig.\ref{figure4}).
The positions $x_n$ ($n=0, 1,...$) of the bands obey the spacing-law, 
i.e., they form a geometric
series $x_n\sim (1+p)^n$ ($p>0$), as observed in experiments and
reproduced by various theories~\cite{MatPack98,ModelB}.

Since the precipitation
is always initiated in the front, the position of the
band $x_n$ and its time of formation $t_n$ are related by the
so-called {\em time-law},
\begin{equation}
x_n=\sqrt{2D_ft_n}\,.
\label{timelaw}
\end{equation}
From this equation one concludes that the positions $x_n$
of the precipitation bands can be regulated
either by modifying the time-law, or by controlling the $t_n$-s.

The first attempts to change the functional form of the time-law
were based on the idea that the reaction takes place between ions
$A^++B^-\to C$, therefore
the motion of the front 
is potentially affected by applying a {\em constant external electric
field}~\cite{Das,Sultan,Lagzi,Unger,Liese-E-front,Liese-E-pattern}.
The results of both the experimental and theoretical
investigations suggest, however, that neither the locality (i),
nor the diffusive nature (ii) of the front are altered significantly
by a constant electric field, i.e., the time-law~(\ref{timelaw})
is practically not modified for the parameter range of interest in the
experiments.
\begin{figure}[t!]
\includegraphics[clip,width=\columnwidth]{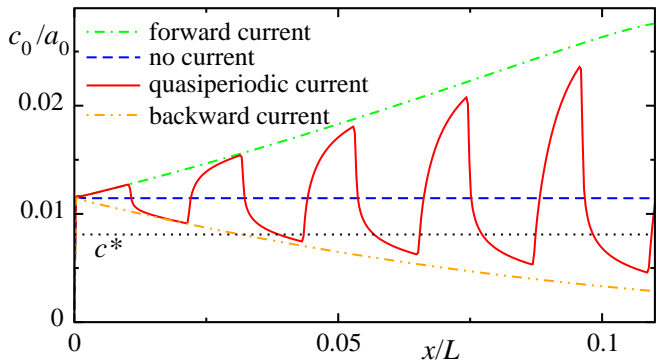}
\caption{Concentrations of the reaction product
in the wake of the front, in the absence of a current, respectively when
a constant forward or backward current,
or a quasiperiodic current (changed at times $\tau n^2/2$)
is switched on.}
\label{figure2}
\end{figure}
The constant electric field has, however, an important effect on 
property (iii) of the front. It was found~\cite{Liese-E-front} that 
for a {\em forward field} (i.e., a field that drives the ionic reagents 
towards each other), the concentration $c$
increases linearly in the direction of the
front motion. A {\it backward field} yields a decreasing
concentration of $C$-s;
this decrease is linear at the beginning and leads finally to the
complete extinction of the reaction. 

The above results equally apply
to the case where a {\em current generator}
is used (Fig.~\ref{figure1}) to produce constant backward and
forward currents, as illustrated in Fig.~\ref{figure2}.
The new results in Fig.~\ref{figure2} (continuous curve) concern the case
when the applied current is constant in absolute value, but changes sign
quasiperiodically at times $\tau n^2/2$, with $\tau$ giving a
timescale and $n=0,1,2,\ldots$.

Figure~\ref{figure2} together with the time-law~(\ref{timelaw})
provide us the key
to create precipitation bands at {\em arbitrarily predefined locations},
by using an appropriately-designed time-dependent current $I(t)$.
Indeed, given a set of prescribed band positions $x_n$, one switches
on the forward current at times $t_n=x_n^2/2D_f$. This
increases the concentration of $C$-s to cross the precipitation
threshold and thus a band forms at $x_n$. In order to avoid
spontaneous formation of spurious bands one must switch on a
backward current at some intermediate times between $t_n$ and $t_{n+1}$.

The above protocol works, in particular, for producing the much
sought after {\it periodic Liesegang pattern}. One can
obtain a controlled wavelength $d$  such that $x_n=d\,n$
by switching on the forward current
at $t_{n}=(2n)^2\tau/2$, where $\tau = d^2/2D_f$.
If the desired period $d$ is smaller than half the
{\it local wavelength} of the
Liesegang pattern, then the spurious bands can be avoided by
switching on the backward current when the front is halfway between
$x_n$ and $x_{n+1}$, i.e., at times
$(2n+1)^2\tau/2$.

In order to put the above argument on a more solid
foundation, we extended our previous model of Liesegang pattern formation
in the presence of an electric field~\cite{Liese-E-front,Liese-E-pattern}
to the case of a time-dependent current flowing through the system.
The first stage of the process is described by the evolution equations for
the concentration profiles of the ions
$a^{\pm}(x,t)$ and $b^{\pm}(x,t)$, with the underlying
electroneutrality hypothesis.
These equations take a relatively simple form for the
case of monovalent ions with equal diffusion coefficients,
\begin{eqnarray}
\partial_t a^+&=&D\partial^2_x a^+- j(t)\,\partial_x(a^+/\Sigma)-ka^+b^-
\label{e-1}\\
\partial_t b^-&=&D\partial^2_x b^-+ j(t)\,\partial_x(b^-/\Sigma)\,-ka^+b^-
\label{e-2}\\
\partial_t a^-&=&D\partial^2_x a^-+ j(t)\,\partial_x(a^-/\Sigma) \label{e-3}\\
\partial_t b^+&=&D\partial^2_x b^+- j(t)\,\partial_x(b^+/\Sigma)
\label{e-4} \, .
\end{eqnarray}
Here $j(t)=I(t)/{\cal A}$ is the externally controlled electric
current-density flowing
through the tube of cross section ${\cal A}$,
$\Sigma=q(a^++a^-+b^++b^-)$ with
$q$ being the unit of charge, and $D$ is the diffusion coefficient
of the ions. The reaction rate $k$ is usually large resulting in a
reaction zone of negligible width.

The second stage in the process, namely
the phase separation of the reaction product $C$,
is modeled by the Cahn-Hilliard equation with a source term
describing the rate of production
of $C$-s in the reaction zone~\cite{ModelB,Liese-E-pattern}.
The free energy underlying the thermodynamics is assumed to have a
Ginzburg-Landau form with minima
at some low ($c_l$) and high ($c_h$) concentrations of $C$.
Using then a shifted and rescaled concentration
$m=(2c-c_h-c_l)/(c_h-c_l)$, yields the following equation
\begin{equation}
\partial_t m=-\lambda\Delta( m - m^3 + \sigma \Delta m)+S(x,t)
\label{cahn-hilliard}
\end{equation}
where $S(x,t)=2ka^+b^-/(c_h-c_l)$ is the source term
coming from the equations (\ref{e-1}--\ref{e-4}) of the first stage.
The parameters $\lambda$ and $\sigma$
can be chosen so as to reproduce the correct time and lengthscales in
experiments~\cite{RZ2000,Liese-E-pattern}.

\begin{figure}[t!]
\includegraphics[width=\columnwidth,clip]{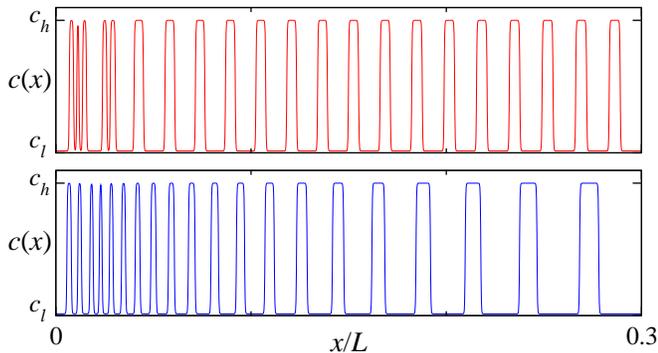}
\caption{Theoretical precipitation patterns.
The periodic pattern emerging in the presence of a
quasiperiodic current (upper panel)
as compared to the usual Liesegang
band-structure obtained in the absence of current (lower panel). The values $c_h$ and $c_l$
are the stable concentrations of  $C$.}
\label{figure3}
\end{figure}

\begin{figure}[t!]
\includegraphics[width=\columnwidth,clip]{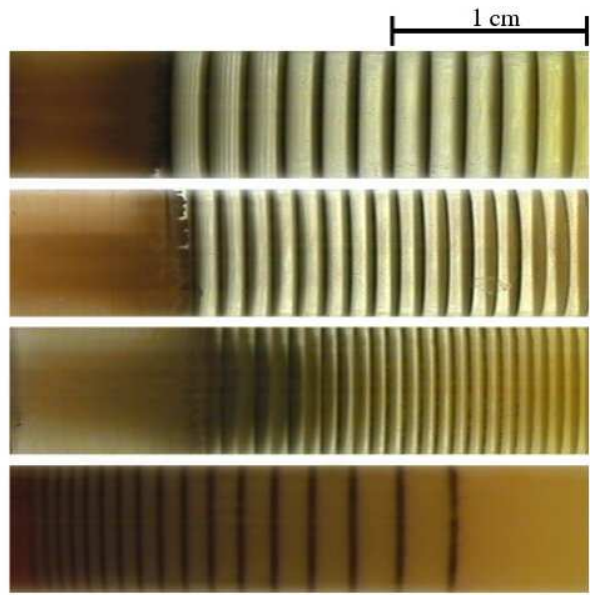}
\caption{Experimental precipitation patterns. A
quasiperiodic current of amplitude 500 $\mu$A was used
with $\tau=4,\,2\,$, and $1$ min, respectively,
as going down the panels.
Lowest panel illustrates the usual Liesegang bands.}
\label{figure4}
\end{figure}

Equations (\ref{e-1}-\ref{cahn-hilliard}) constitute the mathematical
formulation of the problem. They can be solved e.g. by
the classical fourth-order Runge-Kutta method. Examples
of emerging patterns in case of quasiperiodic current (periodic pattern)
or without current (standard Liesegang pattern)
are shown in Fig.~\ref{figure3}~\cite{beginning}.
We tested our theory on a much-studied case where a precipitate of
silver dichromate (${\rm Ag_2Cr_2O_7}$) is formed due to the
reaction of silver nitrate (${\rm AgNO_3}$) and
potassium dichromate (${\rm K_2Cr_2O_7}$) in a gelatine gel.
In this system, various structures have been observed
(regular patterns, spirals, helixes~\cite{muller83}), and the
effect of a constant electric field~\cite{Lagzi} has also been
studied.
The experiments were carried out in vertical glass tubes
(diameter: 1 cm, length: 20 cm)
containing a gelatine gel column (length: 13 cm) closed
by agarose gel stoppers (length: 1.5 cm).
The inner electrolyte was dissolved in the gel
[3.0 g gelatine (Reanal) added to 50~$\rm cm^3$ of 0.0036 mol/$\rm dm^3$
${\rm K_2Cr_2O_7}$ (Reanal) solution] while
the outer electrolyte [2~$\rm cm^3$ of ${\rm AgNO_3}$
solution (Reanal, 2.50 mol/$\rm dm^3$)] was placed on top of the gel
at the start ($t=0$) of the experiments. Ni electrodes were fixed at the top
and the bottom of the gel
and the current was supplied by a
programmable current generator (Keithley 2410 Source Meter).
Initially, the current generator maintained a forward current
(the upper electrode in the ${\rm AgNO_3}$ solution was positive).
The current was then changed from forward to backward and vice versa
every $\tau n^2/ 2$ seconds.
At $t=0$, the outer electrolyte started to diffuse into the gel and,
as the reaction front advanced along the tube,
a brown ${\rm Ag_2Cr_2O_7}$ precipitate emerged in the form of bands.
The experiments were run usually for two days, followed by taking
pictures of the resulting pattern in transmitted light.

\begin{figure}[t!]
\includegraphics[width=\columnwidth,clip]{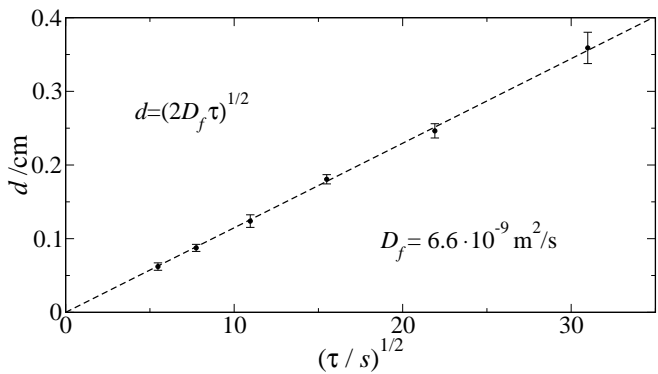}
\caption{Characteristic wavelength $d$ of the pattern generated
by switching  the current forward and backward at times
$\tau n^2/2$. The front diffusion coefficient $D_f$ is the result
of a linear fit.}
\label{figure5}
\end{figure}

Different types of experiments were performed. First,
in the absence of electric current, usual Liesegang bands were produced
(lowest panel of Fig~\ref{figure4}).
Second, a quasiperiodic electric current of a few hundred $\mu$A
amplitude is switched backward and forward at times
$\tau n^2/2$. The resulting periodic patterns for various $\tau$-s
are shown in the three upper panels of Fig.~\ref{figure4}. As can
be seen on Fig.~\ref{figure5},
the wavelengths $d$ of the patterns show the $\sqrt{\tau}$-dependence
in agreement with the theoretical predictions.
It is also remarkable that the wavelengths are
not affected by the intensity of the electric current used in the
experiments (250, 500, and 1000 $\mu$A).
We note that, occasionally, secondary patterns patterns appear
with a wavelength an order of magnitude smaller than $d$
(see the first three band-gaps in the upper panel of
Fig.~\ref{figure4}). It is not quite
clear under which experimental conditions
these secondary patterns form, though they can be observed
more frequently when
$d$ is larger. Our theory does not
account for such features.

\begin{figure}[t!]
\includegraphics[width=1.\columnwidth,height=6.2cm,clip]{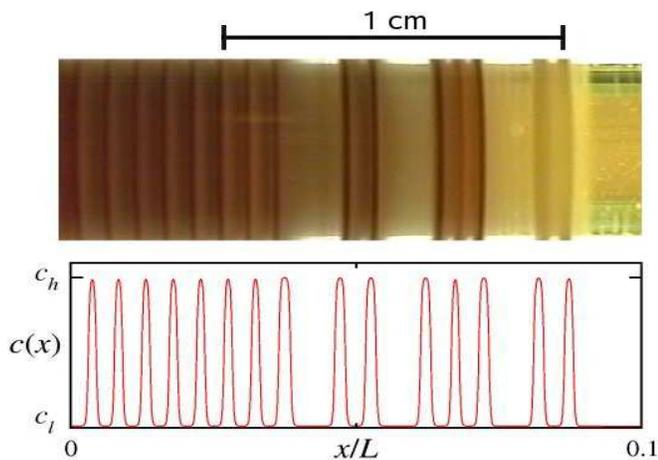}
\caption{An example of designer pattern.
The proposed protocol for generating the "2-3-2" structure is verified
both in the experiment (upper panel) and in the theory (lower panel),
as described in the text.}
\label{figure6}
\end{figure}

A third series of experiments was devoted to demonstrate 
the feasibility of creating more complex
patterns. Figure~\ref{figure6} shows an example where
equidistant bands are followed by a predesigned structure consisting
of groups of 2, 3, and 2 bands separated
by voids. This "2-3-2" pattern  was
generated by taking the protocol for a periodic pattern,
and making two modifications.
Namely, voids are generated by substituting the forward with
a backward current and, furthermore, the
amplitude of the backward current was always
half of the forward one. Here again, the
experimental results are in excellent agreement with the
theoretical predictions (Fig.~\ref{figure6}).
Our results on the control of precipitation patterns
have several important implications.
From a theoretical point of view, they
demonstrate the predictive power and, implicitly, the correctness of our
{\it phase-separation in the
presence of a moving source} basic scenario for Liesegang pattern formation~\cite{ModelB}. 
Indeed, these are the first experiments
which test intricate details of the theory and the agreement
is excellent.

From an experimental point of view, the electric-current control
of the patterns is a flexible and rather general tool, which should be
useful whenever the patterns emerge as the
result of reactions between charged particles
(ions, nano- or colloid particles).
Combining this tool with more traditional ones
(choice of geometry, initial concentrations, boundary conditions)
opens up further possibilities for control and structure design.

One of the main motivation for finding control mechanisms
is to design structures on small scales. This brings up the
question about the lower limit of $d$ obtainable by our method.
Although Fig.\ref{figure5} does not show the existence of such a
limit as $\tau$ is decreased, we should note that there are
at least two effects which are not well-controlled and may limit
the smallest wavelength that can be reached. They are the
width of the reaction zone and the thermal fluctuations
in the system. Their relevance in the downsizing challenge
remains to be understood and evaluated.

\acknowledgments
This work has been supported by the Swiss National Science
Foundation and  by the Hungarian Academy of Sciences (Grants
No.\ OTKA K68109 and K68253). IL also acknowledges support from the
Hungarian Academy of Sciences through the Bolyai Research
Fellowship.


\begin{thebibliography}{99}
\bibitem{beginning} The very beginning of the
pattern is rather irregular (since this is the region where the local
wavelength of the Liesegang pattern is smaller
than $d$, thus our protocol cannot prevent spurious-band
formation).
Later on, the pattern is perfectly periodic.
\end{thebibliography}
\end{document}